\documentclass[
  aps,pra,twocolumn, 
  superscriptaddress,
  nofootinbib      
]{revtex4-2}

\usepackage[a4paper, 
            left=1.5cm,   
            right=1.5cm,  
            top=2.5cm,
            bottom=2.5cm]{geometry}
            
\usepackage[T1]{fontenc}
\usepackage[utf8]{inputenc}     
\usepackage{amsmath}       
\usepackage{amssymb}
\usepackage{mathtools}
\usepackage{graphicx}          
\usepackage{geometry}            
\usepackage{times}
\usepackage{quantikz}

\usepackage[colorlinks=true, linkcolor=blue, urlcolor=blue, citecolor=blue]{hyperref}
\usepackage{titlesec}
\usepackage{float}

\titleformat{\subsubsection}{\normalfont}{\thesubsubsection}{1em}{\itshape}

\begin{document}

\title{Optimizing Quantum State Transformation Under Locality Constraint}

\author{Sasan Sarbishegi}
\affiliation{Department of Physics, Sharif University of Technology,
PO Box 11155-9161, Tehran, Iran}
\author{Maryam Sadat Mirkamali}
\affiliation{Department of Physics, Sharif University of Technology,
PO Box 11155-9161, Tehran, Iran}

\begin{abstract}

In this paper, we present a general numerical framework for both deterministic and probabilistic quantum state transformations, under locality constraints. For a given arbitrary bipartite initial state and a desired bipartite target state, we construct an optimized local quantum channel that transforms the initial state into the target state with high fidelity. To achieve this goal, local quantum channels are parametrized on a complex Stiefel manifold and optimized using gradient-based methods. We demonstrate that this approach significantly enhances entanglement distillation for weakly entangled states via two complementary strategies: optimized local state transformation and probabilistic local transformation. These results establish our method as a powerful and versatile tool for a broad class of quantum information processing tasks.

\end{abstract}

\maketitle

\section{Introduction}

Transforming one quantum state into another under various operational constraints is a fundamental problem in quantum control. Any mixed state can be converted to any arbitrary mixed state in the same Hilbert space by a completely positive trace-preserving (CPTP) map, or a quantum channel \cite{wu2007controllability}. 
However, imposing experimentally motivated restrictions on CPTP maps alters which transformations are achievable \cite{kondra2023real,regula2022tight,chefles2000deterministic,chefles2004existence,spee2024transformations,regula2022probabilistic}.
For example, allowing only real operations \cite{kondra2023real}, requiring deterministic maps between specified finite sets of pure input and output states \cite{chefles2000deterministic,chefles2004existence}, or allowing only finite rounds of local operation and classical communication (LOCC) in distributed networks \cite{spee2024transformations}, substantially shrinks the set of achievable state conversions.
One of the key practical constraints is locality. Consider a bipartite system where each party may be controlled locally. For pure bipartite states, deterministic convertibility using local operations assisted by classical communication (LOCC) is fully characterized by Nielsen’s majorization criterion\footnote{A pure state $|\psi_{\mathrm{in}}\rangle$ can be transformed into $|\psi_{t}\rangle$ iff the Schmidt vector of $|\psi_{\mathrm{in}}\rangle$ majorizes that of $|\psi_{t}\rangle$.}\cite{nielsen1999conditions}. If the majorization condition fails, one can seek a deterministic LOCC transformation that maximizes the fidelity to a target state \cite{vidal2000approximate}. In contrast, for mixed bipartite states under locality constraints, there is no known criterion that guarantees the existence of a local channel mapping an initial state $\rho^{\mathrm{int}}$ to a target state $\rho^{t}$ \cite{chitambar2014everything}, despite the importance of such maps for entanglement distillation \cite{bennett1996purification,deutsch1996quantum}, state discrimination \cite{bennett1999quantum,fan2004distinguishability,walgate2000local,horodecki2003local,nathanson2005distinguishing}, and quantum repeater \cite{inesta2023optimal,goodenough2025noise,li2021efficient}.
                                                                            
In this paper, we address the problem of transforming a given bipartite initial state \(\rho^{\mathrm{int}}_{AB}\) into a state that approximates a target state \(\rho^{t}_{AB}\) as closely as possible, according to a chosen distance metric, by means of local CPTP maps. As the first method, we develop a numerical framework that explores the space of local channels $\Lambda^{\mathrm{loc}}$ to optimize the transformation of $\rho^{\mathrm{int}}_{AB}$ to $\rho^{t}_{AB}$. Concretely, we parameterize $\Lambda^{\mathrm{loc}}$ as points on a complex Stiefel manifold, as in the nonlocal case of Ref.~\citep{oza2009optimization,ahmed2023gradient,russkikh2025quantum}, and for a fixed $\rho^{\mathrm{int}}_{AB}$ and $\rho^{t}_{AB}$ define a differentiable cost function measuring the distance between $\Lambda^{\mathrm{loc}}(\rho^{\mathrm{int}}_{AB})$ and $\rho^{t}_{AB}$.  By minimizing the cost function via a gradient-based algorithm, we numerically identify an optimized local channel \(\Lambda^{\mathrm{loc}}\) such that the resulting state 
\(\Lambda^{\mathrm{loc}}(\rho^{\mathrm{int}}_{AB})\) provides the closest approximation to \(\rho^{t}_{AB}\).

A closely related approach is probabilistic local transformation, where quantum channels are interpreted as local generalized quantum measurements (with associated positive operator valued measures (POVMs)). After performing
the measurement, depending on the measurement outcome,
one may decide whether to retain or discard the resulting post-measurement
state.
Building on this framework, our second method focuses on optimizing the fidelity between the post-selected state and the target state, $\rho^t_{AB}$. 
This optimization is performed by varying the parameters of the local quantum channels, which define the elements of the local generalized measurements, while employing a fixed post-selection rule.

As a key application, we apply both methods to the problem of entanglement distillation.  In general, entanglement distillation consists of
applying local transformation and measurement to noisy bipartite input states,
followed by discarding certain subsystems, with the goal of probabilistically increasing the fidelity of the output entangled state with a maximally entangled state \cite{bennett1996purification}. Numerous works have studied how to optimize output fidelity or success probability under various constraints on both allowed operations and the form of the input state \citep{ruan2018adaptive,krastanov2019optimized,rozpkedek2018optimizing,goodenough2024near,opatrny1999optimization,zhao2021practical,riera2021entanglement,riera2021entanglement2}.  We apply the introduced numerical state-to-state transfer and probabilistic local transformation scheme to the problem of entanglement distillation. The effectiveness of these methods is demonstrated for weakly entangled input states.

In the state-to-state transfer approach, we overcome a limitation of known recurrence protocols, DEJMPS \cite{deutsch1996quantum} and BBPSSW \cite{bennett1996purification}, which act on two noisy entangled pairs and iteratively purify one pair at the expense of the other. These protocols break down when the \emph{fully entangled fraction} (FEF)-the maximal overlap of a state with any maximally entangled state \cite{bennett1996mixed}—falls below \(0.5\).\footnote{Although every 2-qubit state with \(\mathrm{FEF}>0.5\) is necessarily entangled, some weakly entangled states (e.g.\ as measured by concurrence) have \(\mathrm{FEF}\leq0.5\).} Using our optimization framework, we construct a customized distillation protocol for such weakly entangled input states (\(\mathrm{FEF}\leq0.5\)), mapping them as close as possible to the R-states \cite{rozpkedek2018optimizing} that are ideally suited for the extreme photon loss (EPL) protocol \cite{campbell2008measurement}. The transformed states can then be distilled using the EPL protocol. Our numerical results confirm the effectiveness of this tailored strategy in distilling states with low fully entangled fraction.

In the probabilistic local transformation approach, rather than optimizing a transformation followed by a specific distillation protocol, we directly optimize a single-step protocol aimed at mapping an arbitrary 2-qubit input state to a Bell state (e.g. \(\lvert\Psi^{+}\rangle\langle\Psi^{+}\rvert\)) by selecting suitable local generalized measurement elements together with a post-selection rule that determines whether the output is kept or discarded. In other words, we employ numerically optimized probabilistic local transformation to increase fidelity with respect to \(\lvert\Psi^{+}\rangle\langle\Psi^{+}\rvert\). This procedure parallels the scheme proposed in Ref.~\cite{gisin1996hidden}, known as the \emph{filtering protocol}; however, in our case, the operators are specifically tailored to the known input state to maximize its fidelity with the target Bell state. We show that this approach provides optimal local operations for distillation by comparing the results for 50 input real noisy entangled states with the theoretical upper bound of Ref.~\cite{rozpkedek2018optimizing}. For all 50 random samples, our approach saturates the upper bound. Next, we perform both approaches on 50 general (not restricted to real) random weakly entangled states and compare the results. We show that the second approach is more effective in distilling weakly entangled states with low FEF.

Beside entanglement distillation, the two introduced approaches may be applied to any state transformation problem with locality constraint (and known initial state).  

\section{Statement of the Problem}
To tackle the problem of transforming a bipartite mixed state \(\rho^{\mathrm{int}}_{AB}\) into a target state \(\rho^t_{AB}\) using a local channel \(\Lambda^{\mathrm{loc}}\), we use the Kraus representation of CPTP maps. Any quantum channel \(\Lambda\) admits a decomposition
\(
\Lambda(\rho) = \sum_i K_i\,\rho\,K_i^\dagger,
\)
where the Kraus operators $\{ K_i \}$ satisfy the trace-preserving relation $\sum_i K_i^\dagger K_i = \mathbb{I}$. A local channel \(\Lambda^{\mathrm{loc}}\) is specified by tensor product Kraus operators \(\{K_i^{(A)} \otimes K^{(B)}_j\}\), where Kraus maps \(\{K_i^{(A)}\}\) and \(\{K_j^{(B)}\}\) act on local Hilbert spaces \(\mathcal{H}_A\) and \(\mathcal{H}_B\), respectively.\footnote{At this stage, we do not include classical communication in our operations.} The action of \(\Lambda^{\mathrm{loc}}\) on a bipartite state \(\rho^{\mathrm{int}}_{AB}\) is given by
\begin{align}  
\Lambda^{\mathrm{loc}}(\rho^{\mathrm{int}}_{AB})
= \sum_{i,j} \bigl(K_i^{(A)} \otimes K^{(B)}_j\bigr)\,
\rho^{\mathrm{int}}_{AB}\,
\bigl(K_i^{(A)} \otimes K^{(B)}_j\bigr)^\dagger.
\label{eq:local kraus representation} 
\end{align}
The goal is to find the set of Kraus operators \(\{K_i^{(A)}\}\) and \(\{K_j^{(B)}\}\) such that \(\Lambda^{\mathrm{loc}}(\rho^{\mathrm{int}}_{AB})\) is close to \(\rho^t_{AB}\).

To quantify how well \(\Lambda^{\mathrm{loc}}(\rho^{\mathrm{int}}_{AB})\) approximates \(\rho^t_{AB}\), we introduce a distance function \(D(\rho_1,\rho_2)\) between two arbitrary states. This metric is defined via the differences in expectation values of a complete set of independent observables that uniquely determine an \(N\) dimensional density matrix:
\begin{align}
D(\rho_1, \rho_2\ ; M_i) \coloneqq \sqrt{  \sum_{i=1}^{N^2-1} \left( \text{Tr} \left[ \rho_1 M_i \right] - \text{Tr} \left[ \rho_2 M_i \right] \right)^2 } \label{metric}
\end{align}
where \(\{ M_i \}\) is a set of \(N^2-1\) Hermitian normalized measurement operators representing the chosen independent observables.\footnote{ \(N^2-1\) is the number of measurements needed to characterize a density matrix defined on a Hilbert space of dimension \(N\).} Each measurement operator follows the normalization relation, \(\text{Tr}\left[M_i^\dagger M_i\right]=1\). This distance is zero only if \(\rho_1 = \rho_2\), as the expected values of the measurements coincide. Accordingly, our optimization task reduces to finding \(
\min_{\Lambda^{\mathrm{loc}}}\ D\bigl(\Lambda^{\mathrm{loc}}(\rho^{\mathrm{int}}_{AB}),\,\rho^t_{AB}\ ;M_i\bigr),
\)
where minimization runs over all local CPTP maps \(\Lambda^{\mathrm{loc}}\). We choose this metric over other conventional options such as trace distance or fidelity because the optimization framework (detailed in the next section) is tailored to maximize linear functionals of the form \(\ \text{Tr}\bigl[H\,\rho\bigr]\). These functionals depend linearly on the density matrix \(\rho\), with \(H\) being a Hermitian operator.

Following the construction in Ref.~\citep{oza2009optimization}, for any map \(\Lambda\) the associated Kraus operators \(\{K_i\}_{i=1}^{N^2}\) can be assembled into a single block matrix \(S\):
\begin{align}
S = \begin{pmatrix}
K_1 \\
K_2 \\
\vdots \\
K_{N^2}
\end{pmatrix},
\label{eq:stack_S}
\end{align}
By construction, the trace-preserving condition on the Kraus operators imposes \(S^\dagger S = \mathbb{I}_{N}\) on the matrix \(S\). The stacked matrix \(S\) lies on a complex Stiefel manifold defined below. 

Given field \(\mathbb{F}\) (in our setting \(\mathbb{F}=\mathbb{C}\)), the Stiefel manifold \(V_{l}(\mathbb{F}^{n})\) is the set of all ordered orthonormal \(l\)-frames in \(\mathbb{F}^{n}\)~\citep{boothby2003introduction}; Alternatively, by representing each \(l\)-frame as a $n\times l$ matrix $S$, a complex Stiefel manifold can be characterized as
\begin{align}
V_{l}(\mathbb{C}^{n}) = \left\{ S \in \mathbb{C}^{n \times l} \mid S^\dagger S = \mathbb{I}_{l} \right\}. \label{Stiefel_definition}
\end{align}

Therefore, finding a local channel that transforms \(\rho^{\mathrm{int}}_{AB}\) to \(\rho^t_{AB}\), reduces to defining a differentiable cost function. We choose the distance function defined in Eq.~\eqref{metric}, between \(\Lambda^{\mathrm{loc}}(\rho^{\mathrm{int}}_{AB})\) and \(\rho^t_{AB}\) as the cost function:
\begin{align}
f\bigl(S;\,\rho^{\mathrm{int}}_{AB},\rho^t_{AB},M_i\bigr) \coloneqq 
D\bigl(\Lambda^{\mathrm{loc}}(\rho^{\mathrm{int}}_{AB}),\,\rho^t_{AB}\ ;M_i\bigr)
\end{align}
By minimizing this cost function over the Stiefel manifold, we identify the optimal point on the Stiefel manifold, \(S_{\mathrm{op}}\), corresponding to the optimal local map \(\{K_i^{(A)} \otimes K^{(B)}_j\}_{\mathrm{op}}\) that best approximates the transformation to the target state. We employ a gradient descent algorithm to efficiently locate the minimum of the differentiable cost function \(f\). The minimum reached by this procedure is possibly a local minimum for the chosen objective function. Therefore, it is best to repeat the optimization process with multiple initial states.

With known Kraus operators corresponding to the desired channel, the channel can be experimentally implemented by coupling the system to an ancilla and applying a joint unitary operation.  Note that for a system defined on an \(N\)-dimensional Hilbert space, one can use an ancilla of dimension \(N^{2}\), prepare it in an initial pure state, and enact a global unitary \(U_{\text{sys-anc}}\) whose action reproduces the Kraus map on the system when the ancilla is subsequently traced out.

A quantum channel can be regarded as a generalized measurement, performed
without post-selection. By applying a selection strategy to the measurement outcomes, one obtains a probabilistic operation acting on a quantum state. This approach is discussed as the second method in this paper and it is named probabilistic local transformation.

The rest of the paper is organized as follows. In Section~\ref{3}, after briefly reviewing the parameterization of quantum channels on a complex Stiefel manifold, we propose a novel scheme for imposing locality constraints on channel operators, and describe how to minimize the resulting cost function over this restricted set of local channels. In Section~\ref{4}, we show a key application of our framework to the optimization of entanglement distillation for weakly entangled states, comparing two distinct optimization strategies and providing corresponding numerical results. Finally, in Section~\ref{5}, we summarize our findings and discuss their implications and potential future directions.   

\section{Optimization Procedure} \label{3}
In this section, we present the underlying principles of the gradient-based optimization method for optimizing local CPTP maps via Stiefel manifold optimization. We begin by reviewing how to parameterize Kraus operators as points on a complex Stiefel manifold, following the approach of Ref.~\citep{oza2009optimization}. Then, we address the optimization problem of maximizing the expectation value of a specified observable over CPTP maps that evolve a given initial state. Next, we impose a locality constraint that requires that all Kraus operators remain separable. Building upon this constraint, we formulate the bipartite state transformation task under a local CPTP map and apply our gradient‑based algorithm to identify the local channel that most closely approximates the desired transformation.  Finally, we extend the framework to probabilistic local transformation by incorporating classical communication. We treat the local operation as a generalized quantum measurement and use the communicated measurement outcomes to decide whether to keep or discard each resulting state.

\subsection{Kraus Operators and the Stiefel Manifold: A Review}
As a preliminary step, in this section we address the following question: Given an initial state \(\rho^{\mathrm{int}}\) and an observable \(\mathcal{O}\), what is the optimal CPTP map that evolves \(\rho^{\mathrm{int}}\) into a state \(\rho^t\) while maximizing the expectation value of the observable \(\mathcal{O}\). There are yet no restrictions on the CPTP map. In the next section, we constrain the CPTP map to local ones. We answer this question based on a gradient-based numerical approach given in Ref.~\cite{oza2009optimization}. 

Every CPTP map $\Lambda: \mathcal{L}(\mathcal{H_A)} \to \mathcal{L}(\mathcal{H_A)}$, from the linear operators on Hilbert space $\mathcal{H}_A$ to itself, can be expressed as
\begin{align}
\Lambda(\rho) = \sum_{i=1}^{N_A^2} K_i \rho K_i^\dagger, \label{eq:kraus_sum}
\end{align}
using a set of at most $N_A^2$ Kraus operators $\{K_i\}$ that satisfies the trace-preserving relation $\sum_{i=1}^{N_A^2}K_i^\dagger K_i=\mathbb{I}_{N_A}$, where $N_A$ is the dimension of $\mathcal{H}_A$ \cite{choi1975completely} . As demonstrated in Eq.~\eqref{eq:stack_S}, stacking these Kraus operators into a matrix \(S\) yields the orthogonality condition \(S^\dagger S = \mathbb{I}_{N_A}\), which follows directly from the given trace-preserving relation. Therefore, the matrix $S$ lies on a complex Stiefel manifold, defined in \eqref{Stiefel_definition}, that is, the set of all \(n_{A}\times l_{A}\) matrices whose columns form a set of orthonormal vectors in \(\mathbb{C}^{n_{A}}\), where $n_A = N_A^3$ and $l_A = N_{A}$. By mapping the Kraus operators to the Stiefel manifold, the expected value of an observable $\mathcal{O}$ can be written as a function $J(S)$ defined on the Stiefel manifold:
\begin{align}
J(S\ ; \mathcal{O},\rho) \coloneqq \langle \mathcal{O} \rangle_{\Lambda(\rho)} = \text{Tr}\left[ S \rho S^\dagger \tilde{\mathcal{O}} \right], \label{expected_value}
\end{align} 
where \(\tilde{\mathcal{O}}\) is defined as
\begin{align}
\tilde{\mathcal{O}}  \coloneqq \mathbb{I}_{N_A^2} \otimes \mathcal{O}. \label{tilde_operators}   
\end{align}
We maximize this function by computing its gradient and applying gradient-based optimization, such as gradient ascent. Using this approach, one will find a CPTP map that transforms a given initial state \(\rho^{{\mathrm{int}}}\) to a final state \(\rho^t\) that maximizes the expectation value of the observable $\mathcal{O}$.

Since the Stiefel manifold is an embedded sub-manifold of the Euclidean space $\mathbb{C}^{n_A \times l_A}$, we first compute the gradient in the ambient space by Eq.\eqref{Gradient_G} below, and then project it onto the tangent space of the Stiefel manifold, as illustrated in Fig.~\ref{fig:manifold depiction}.
\begin{align}
G(S) \coloneqq \nabla J(S) = 2 \tilde{\mathcal{O}} S \rho, \label{Gradient_G}
\end{align}
The gradient vector $G$ at the point $S$ is then projected onto the tangent space at \(S\), denoted by $T_S V_{l_A}(\mathbb{C}^{n_A})$, where \(V_{l_A}(\mathbb{C}^{n_A})\) is the Stiefel manifold defined in Eq.~\eqref{Stiefel_definition}, and \(T_S\) stands for tangent space at \(S\). This is achieved using the linear projection below, which maps any vector $X \in \mathbb{C}^{n_A \times l_A}$ to the tangent space $T_S V_{l_A}(\mathbb{C}^{n_A})$:
\begin{align}
\pi_{S}(X) &= X-\frac{1}{2}S(S^\dagger X+X^\dagger S).
\label{projection_nonlocal}
\end{align}
Once the gradient of $J(S)$ is obtained, one can use the Cayley transformation and curvilinear search as an optimization algorithm \citep{wen2013feasible}. However, as applied in Ref.~\cite{oza2009optimization}, for simplicity of implementation and reduced computational overhead, we adopt a first–order gradient ascent method on the manifold. Starting from a randomly chosen initial point \(S_{0}\) on the Stiefel manifold, we iteratively move in the direction of the projected gradient, thereby generating a sequence \(\{S_{t}\}\) according to
\begin{align}
  S_{t+1}
  \;=\;
  S_{t}
  \;+\;
  \alpha_{t}\,
  \pi_{S_{t}}\!\bigl(G(S_{t})\bigr),
\end{align}
where $\alpha_t > 0$ is an appropriately chosen step size, and \(t\) indexes the iteration.\footnote{Throughout this paper we fix the step size and set \(\alpha_{t}=\alpha\) for all iterations.} Note that, for a minimization task, gradient descent can be applied by moving in the direction opposite to the gradient.

The same framework may accommodate searches over specialized channels. For example, to consider CPTP maps with low Kraus rank, one simply adjusts the dimensions of the matrix \(S\) introduced in Eq.~\eqref{eq:stack_S}. Furthermore, as described in the following subsection, the framework can be used to restrict the channels to local channels.

\subsection{Locality Constraint on Kraus Operators}

In the remainder of Section~\ref{3}, we provide the mathematical framework underlying the main contribution of this paper. We present a numerical approach to determine the optimal local transformation that evolves a given bipartite initial state into a desired final state. In particular, we address the problem of imposing a locality constraint on the CPTP map, which in turn requires the Kraus operators to be separable.
Suppose that we are asked to find the maximum expected value of the observable \(\mathcal{O}\) when the operations are constrained to be local (for now, we ignore the classical communication). In other words, we need to find a set of Kraus operators \(\{K_i^{(A)} \otimes K^{(B)}_j\}\) associated with a local CPTP map \(\Lambda^{\text{loc}}:\mathcal{L}(\mathcal{H}_A\otimes\mathcal{H}_B)\to \mathcal{L}(\mathcal{H}_A\otimes\mathcal{H}_B)\). The set \(\{K^{(A)}_i\}\) acts only on the subsystem \(A\) and \(\{K^{(B)}_j\}\) only on \(B\), so that \(\text{Tr}[\Lambda^{\text{loc}}(\rho_{AB}) \, \mathcal{O}] \) is maximized. Note that the action of this map on a bipartite state \(\rho_{AB}\) is given by
\begin{align}
\Lambda^{\text{loc}}(\rho_{AB}) = \sum_{i,j=1}^{N_A^2,N_B^2 } (K_i^{(A)} \otimes K^{(B)}_j) \rho_{AB} (K_i^{(A)} \otimes K^{(B)}_j)^\dagger,
\end{align}

\begin{figure}
    \centering
    \includegraphics[width=0.7\linewidth]{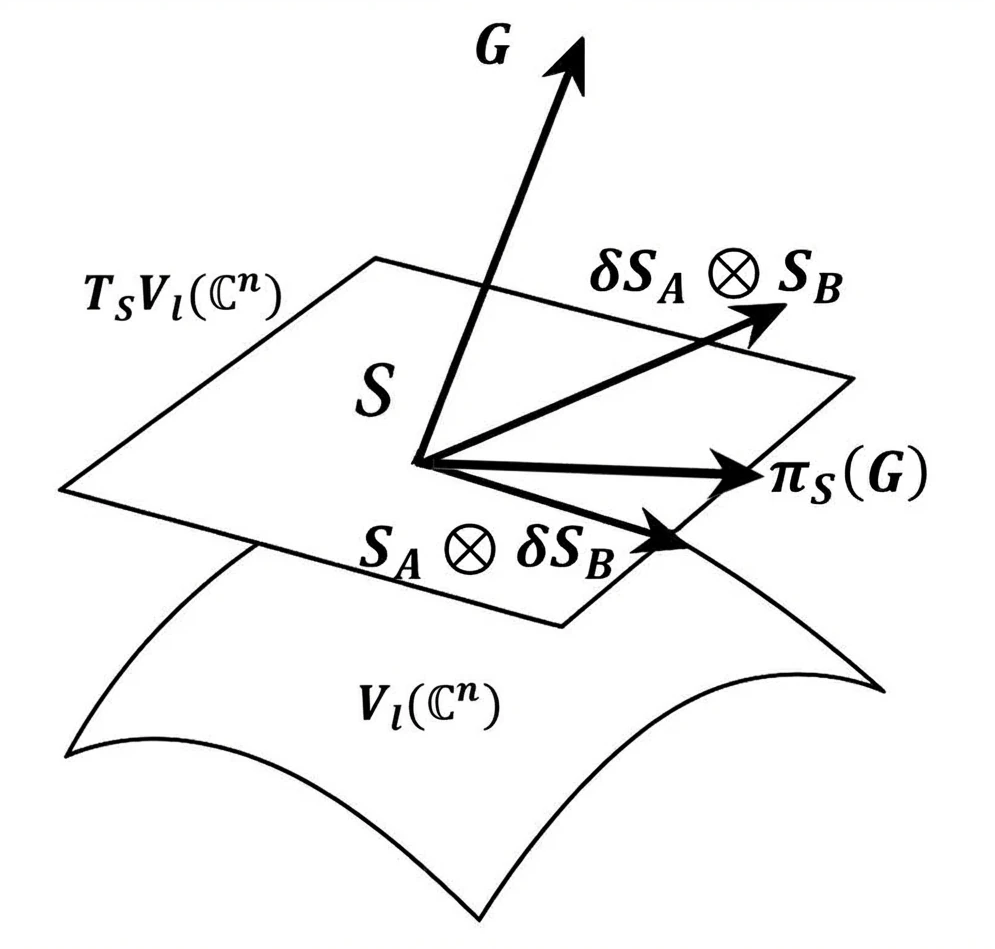}
    \caption{After evaluating the gradient vector \(G\) at the point \(S\), we project it onto one of two subspace coordinates of the tangent space \(T_S V_k(\mathbb{C}^n)\), either \(S_A \otimes \delta S_B\) or \(\delta S_A \otimes S_B\). These subspaces allow variation in only one subsystem's parameters, effectively constraining the optimization to local updates.}
    \label{fig:manifold depiction}
\end{figure}

As in the nonlocal case, we can represent all Kraus operators in a single matrix \(S\), which lies on a Stiefel manifold. However, in this case, \(S\) can be constructed from matrices \(S_A\) and \(S_B\), which correspond to the Kraus operators \(\{K^{(A)}_i\}\) and \(\{K^{(B)}_j\}\), respectively. Specifically, by permuting the rows of the tensor product \(S_A \otimes S_B\), we obtain the matrix \(S\):
\begin{align}   
S = U_\sigma (S_A \otimes S_B) = \begin{pmatrix}
K_1^{(A)} \otimes K^{(B)}_1 \\
K_1^{(A)} \otimes K^{(B)}_2 \\
\vdots \\
K_{N_A^2}^{(A)} \otimes K^{(B)}_{N_B^2}
\end{pmatrix},
\end{align}
Here, \(U_{\sigma}\) denotes the unitary matrix that implements the row permutation on \(S_A \otimes S_B\), where \(\sigma\in\mathcal{P}_{n_{A}n_{B}}\) is an element of the permutation group on \(n_{A} n_{B}\) symbols. As stated earlier, \(n_{A}=N_{A}^{3}\) and similarly \(n_{B}=N_{B}^{3}\).  For convenience, we introduce
\begin{align}
\tilde{S}\;\coloneqq\;U_{\sigma}^{\dagger}\,S =S_A\otimes S_B
\end{align}
so that the maximization of the expected value in~\eqref{expected_value} can be rewritten entirely in terms of \(\tilde{S}\).  This reformulation enables us to search directly for the optimal Kraus operators acting on the joint system \(AB\):
\begin{align}
J_{\text{loc} }(\tilde{S} \ ; \mathcal{O},\rho_{AB}) &\coloneqq \text{Tr}\left[ S \rho_{AB} \, S^\dagger \, \tilde{\mathcal{O}} \right] \nonumber \\
&= \text{Tr}\left[ \tilde{S} \rho_{AB} \, \tilde{S}^\dagger \, U_\sigma^\dagger \tilde{\mathcal{O}} U_\sigma \right]. \label{expected_value_local}
\end{align}
Thus, similar to ~\eqref{Gradient_G}, the gradient of the local cost function is  
\begin{align}
  G_{\mathrm{loc}}(\tilde{S})
  \;\coloneqq\;
  \nabla J_{\mathrm{loc}}(\tilde{S})
  \;=\;
2\,(U_{\sigma}^{\dagger}\,\tilde{\mathcal{O}}\,U_{\sigma})\,\tilde{S}\,\rho_{AB}.
\end{align}
To perform a gradient step on the manifold, we must project this vector onto the tangent space at \(\tilde{S}\). 
Since the updated \(\tilde{S}\) must remain separable, the projection must land in the subspace of the tangent space that preserves separability. To impose this restriction, we project \(G_{\mathrm{loc}}(\tilde{S})\) onto the two subspaces
\(S_{A}\otimes T_{S_{B}}V_{l_{B}}\!\bigl(\mathbb{C}^{n_{B}}\bigr)
\quad\text{and}\quad
T_{S_{A}}V_{l_{A}}\!\bigl(\mathbb{C}^{n_{A}}\bigr)\otimes S_{B}\) sequentially. We denote these subspaces as 
\(S_{A} \otimes \delta S_{B}\) and \(\delta S_{A} \otimes S_{B}\), respectively (see Fig.~\ref{fig:manifold depiction}).   
This projection procedure allows us to iteratively update the parameters of subsystem \(B\) while keeping \(A\) fixed, and vice versa.\footnote{Unlike the seesaw method, where one fixes an entire variable, fully optimizes the other, and then swaps roles, our procedure alternates the direction of each infinitesimal step between the two subspaces, so that we modify one subsystem’s parameters only slightly while "momentarily" keeping the other’s fixed.}

To construct these projections, for example, the projection of 
\(\,G_{\mathrm{loc}}(\tilde S)\) onto 
\(S_{A}\otimes \delta S_B \), we begin by projecting the gradient vector onto the ambient space 
\(
S_{A}\otimes \mathbb{C}^{n_{B}\times l_{B}},
\)
where \(\mathbb{C}^{n_{B}\times N_{B}}\) denotes the space of all \(n_{B}\times N_{B}\) complex matrices.  Since this space admits an orthonormal basis
\(\{e_{i}\}\), we may form a corresponding basis for the tensor product as
\(\{S_{A}\otimes e_{i}\}_{i=1}^{2n_Bl_B}\). Hence, the projection is given by
\begin{align}
P_{A}\bigl(G_{\mathrm{loc}}(\tilde S)\bigr)
\coloneqq  \sum_{i=1}^{2n_Bl_B}
   \frac{\bigl\langle G_{\mathrm{loc}}(\tilde S),\,S_{A}\otimes e_{i}\bigr\rangle}
        {\|S_{A}\|^{2}}
   \,\bigl(S_{A}\otimes e_{i}\bigr), \label{projection_local_1}
\end{align}
where the inner product is defined as \footnote{To derive the gradient vector, we adopt the inner product defined by the real part of the Hilbert–Schmidt inner product. This choice requires using a real scalar field, which in turn doubles the number of basis vectors. Concretely, if \(\{e'_j\}_{j=1}^{n_B l_B}\) is a basis for the real vector space \(\mathbb{R}^{n_B \times l_B}\), then the corresponding basis for the complex space \(\mathbb{C}^{n_B \times l_B}\) is
\(
\{e_j\}
= 
\{e'_j\}_{j=1}^{n_B l_B}
\cup
\{\mathrm{i}\, e'_j\}_{j=1}^{n_B l_B},
\)
where \(\mathrm{i}\) denotes the imaginary unit.}

\begin{align}
\bigl\langle G_{\mathrm{loc}}(\tilde S) , S_{A}\otimes e_{i} \bigr\rangle
\coloneqq  \Re  \bigl(\text{Tr}\bigl[G_{\mathrm{loc}}(\tilde S) ^\dagger(S_{A}\otimes e_{i})\bigr]\bigr).
\end{align}
This construction yields a projected gradient of the form
\(
P_{A}\bigl(G_{\mathrm{loc}}(\tilde S)\bigr)
= S_{A}\otimes M_{B}
\), where \(M_B\) is 
\begin{align}
M_B=\frac{1}{\|S_{A}\|^{2}} \sum_{i=1}^{2n_Bl_B}
   \bigl\langle G_{\mathrm{loc}}(\tilde S),\,S_{A}\otimes e_{i}\bigr\rangle
   \,e_{i}.
\end{align}
The remaining task is to project the factor \(M_{B}\) into the tangent space 
\(T_{S_{B}}V_{l_{B}}\!\bigl(\mathbb{C}^{n_{B}}\bigr)\) by applying the projection operator given in Eq.~\eqref{projection_nonlocal}:
\begin{align}
\Pi_{B}\big( P_{A}\bigl(G_{\mathrm{loc}}(\tilde S)\bigr) \big) \coloneqq S_A \otimes \pi_{S_B}(M_B) \label{projection_local_2}
\end{align}

Finally, the same construction can be applied to the other subsystem. Concretely, the gradient
\(
G_{\mathrm{loc}}(\tilde S)
\) is projected onto the tensor‐product subspace \(
T_{S_{A}}V_{l_{A}}\!\bigl(\mathbb{C}^{n_{A}}\bigr)
\;\otimes\;
S_{B},
\) 
thereby allowing the variations of the Kraus operators’ parameters on subsystem \(A\), while keeping the parameters associated with the subsystem \(B\) fixed. In summary, we employ an iterative scheme in which, at each step, we move along two directions:
\(
\Pi_{A}\!\bigl(P_{B}\bigl(G_{\mathrm{loc}}(\tilde S)\bigr)\bigr)
\) and \(
\Pi_{B}\!\bigl(P_{A}\bigl(G_{\mathrm{loc}}(\tilde S)\bigr)\bigr)
\).
By alternately applying these projected gradients, we effectively adjust the Kraus operator parameters for each subsystem in order to maximize the objective function in Eq.~\eqref{expected_value_local}:
\begin{align}
\tilde{S}_{2t+1} &= \tilde{S}_{2t}+\alpha\ \Pi_{A}\!\bigl(P_{B}\bigl(G_{\mathrm{loc}}(\tilde S_{2t})\bigr)\bigr) \\
\tilde{S}_{2t+2} &= \tilde{S}_{2t+1}+\beta \ \Pi_{B}\!\bigl(P_{A}\bigl(G_{\mathrm{loc}}(\tilde S_{2t+1})\bigr)\bigr). \label{evolution_local}
\end{align}

Before we get into real applications in entanglement distillation, we perform two distinct examples that have analytic answers to find out how efficient and accurate the model is working. 
As a first simple and illustrative example, consider the 2‐qubit observable 
\(\mathcal{O} =|00\rangle\langle00| \)
and an arbitrary 2‐qubit state \(\rho^{\mathrm{int}}_{AB}\).  One can always choose local Kraus operators
\(
\bigl\{\,K_{ij} = (|0\rangle\langle i\rvert\,U_a) \otimes (| 0\rangle\langle j\rvert U_b)\bigr\},
\)
with $i,j\in\{0,1\}$ and arbitrary unitaries \(U_a, U_b\in\mathcal{U}(2)\), so that the channel
\(\Lambda(\rho^{\mathrm{int}}_{AB})=\sum_{i,j}K_{ij}\,\rho^{\mathrm{int}}_{AB}\,K_{ij}^\dagger\) maps any input \(\rho\) to \(\lvert00\rangle\langle00\rvert\). Consequently, the maximum possible expectation value can always be achieved. 

Using the presented optimization scheme, we identify the set of Kraus operators that satisfy the locality constraints and maximize the expected value of presented observable for any given 2-qubit density matrix.
We generate a random \(4{\times}4\) density matrix \(\rho_{\mathrm{rand}}\) and apply the
gradient-based optimization method introduced in this section.
Starting from a random point \(S_{A}^{(0)}\otimes
S_{B}^{(0)}\) on the product complex Stiefel manifold,
we employ projected gradient ascent with step size
\(\alpha=\beta=10^{-3}\) and terminate iterations once the variation of the objective function falls below \(10^{-7}\). 
The optimized channel associated with the end point
\(S_{A}^{(t_{f})}\otimes S_{B}^{(t_{f})}\) on the manifold is then tested on
\(100\) independently drawn random states. After transforming these states using the optimized channel and computing the expectation value of \(\lvert00\rangle\langle00\rvert\), the resulting sample mean is
\((1 - 10^{-6}) \pm 10^{-6}\), which is very close to the maximum expected value.

As a second example, we seek to increase the fully entangled fraction (FEF) of a given 2-qubit state. The FEF of a bipartite state \(\rho_{AB}\) is defined as the maximum fidelity between \(\rho_{AB}\) and any maximally entangled state. In the case of two qubits, this quantity can be written as:
\begin{align}
\mathrm{FEF}(\rho_{AB}) &= \max_{|\Phi\rangle }\ F(\rho_{AB},|\Phi\rangle \langle\Phi|) \nonumber\\
&= \max_{|\Phi\rangle } \ \langle \Phi|\rho_{AB}|\Phi\rangle,
\end{align}
where \(|\Phi\rangle\) denotes a maximally entangled state. Equivalently, this corresponds to maximizing the fidelity between \(\rho_{AB}\) and \(\lvert\Phi^+\rangle\) under all local unitaries: 
\begin{equation}
  \label{eq:fef-definition}
  \mathrm{FEF}(\rho_{AB})=
  \max_{U_{a},U_{b}\in\mathcal{U}(2)}
  \langle{\Phi^{+}}|\
    \bigl(U_{a}^\dagger\otimes U_{b}^\dagger\bigr)\,\rho_{AB}\,
    \bigl(U_{a}\otimes U_{b}\bigr)
  \ |{\Phi^{+}}\rangle,
\end{equation}
where \(|\Phi^{+}\rangle=(|11\rangle+|00\rangle)/\sqrt{2}\) is a Bell state. \(U_{a}\otimes U_{b}\) is a local unitary transformation that can produce any maximally entangled state \(|\Phi\rangle\) out of \(|\Phi^+\rangle\) by choosing the appropriate \(U_{a}\otimes U_{b}\). 

In Ref.~\cite{badziag2000local}, it was shown that for density matrices of a particular form (we denote this set by \(\mathcal{C}\)), applying one-sided amplitude damping channel, \(\Lambda_{\rm amp}^{\rm loc} \), can increase the FEF.  To compare, we take a test state
\begin{align}
\rho^\ast =\frac{1}{2} \begin{pmatrix}
0 & 0 & 0 & 0 \\
0 & 3-2\sqrt{2} & 1-\sqrt{2} & 0 \\
0 & 1-\sqrt{2} & 1 & 0 \\
0 &  0 & 0 & 2\sqrt{2}-2
\end{pmatrix},   
\end{align}
which satisfies \(\rho^\ast \in \mathcal{C}\) and has a fully entangled fraction
\(\mathrm{FEF}(\rho^\ast) = 0.5\). We perform gradient ascent with observable \(\mathcal{O}=|\Phi^+\rangle\langle\Phi^+|\) and step size \(\alpha=\beta=10^{-3}\), to obtain a local CPTP map \(\Lambda^{f}\), associated to the end point \(S_{A}^{(t_{f})}\otimes S_{B}^{(t_{f})}\), that maximizes \(\mathrm{FEF}\bigl(\Lambda^{f}(\rho^\ast)\bigr)\). The final value,
\(\mathrm{FEF}\bigl(\Lambda^{f}(\rho^\ast)\bigr)\approx 0.522408\), is
in close agreement with the one-sided amplitude damping result
\(\mathrm{FEF}\bigl(\Lambda^{\rm loc}_{\rm amp}(\rho^\ast)\bigr)\approx 0.522407\) reported in \cite{badziag2000local} to the accuracy $10^{-6}$. The resulting channels are represented by the Kraus sets \(\{K^{(A)}_i\}\) and \(\{K_j^{(B)}\}\), shown in Table~\ref{tab:SA_SB}.

\begin{table}[h]
    \centering
    \caption{Kraus operators of subsystem \(A\) and \(B\)}
    \renewcommand{\arraystretch}{1.4}
    \begin{tabular}{|c|c|}
        \hline
        \textbf{$\{K_i^{(A)}\}$} & \textbf{$\{K_j^{(B)}\}$} \\
        \hline
        $\begin{pmatrix}
            +0.99{+}0.07j & -0.00{+}0.00j \\
            +0.00{+}0.00j & +0.99{+}0.08j
        \end{pmatrix}$
        &
        $\begin{pmatrix}
            -0.00{-}0.00j & -0.04{+}0.07j \\
            +0.06{-}0.10j & +0.73{+}0.07j
        \end{pmatrix}$ \\
        \hline
        $\begin{pmatrix}
            +0.06{-}0.08j & +0.00{+}0.00j \\
            +0.00{-}0.00j & +0.06{-}0.07j
        \end{pmatrix}$
        &
        $\begin{pmatrix}
            +0.00{-}0.00j & -0.27{-}0.50j \\
            +0.41{+}0.75j & +0.08{-}0.09j
        \end{pmatrix}$ \\
        \hline
        $\begin{pmatrix}
            +0.05{+}0.04j & -0.00{+}0.00j \\
            +0.00{+}0.00j & +0.05{+}0.04j
        \end{pmatrix}$
        &
        $\begin{pmatrix}
            -0.00{-}0.00j & -0.10{+}0.21j \\
            +0.15{-}0.31j & +0.04{+}0.05j
        \end{pmatrix}$ \\
        \hline
        $\begin{pmatrix}
            -0.01{+}0.03j & -0.00{-}0.00j \\
            -0.00{-}0.00j & -0.01{+}0.02j
        \end{pmatrix}$
        &
        $\begin{pmatrix}
            -0.00{+}0.00j & +0.07{+}0.23j \\
            -0.11{-}0.34j & -0.01{-}0.02j
        \end{pmatrix}$ \\
        \hline
    \end{tabular}
    \label{tab:SA_SB}
\end{table}

In the following sections, we introduce additional functions built upon the primary objective function in Eq.~\eqref{expected_value_local}. Because the original functional is non-convex, all derived cost functions remain non-convex, necessitating multiple random start points to reduce the risk of converging to local optimum.

\subsection{State Transformation}
By introducing the distance function defined in Eq.~\eqref{metric}, the objective becomes to bring the initial state \(\rho^{\mathrm{int}}_{AB}\) as close as possible to the target state \(\rho^{t}_{AB}\) under a local CPTP map \(\Lambda^{\mathrm{loc}}\).  Equivalently, we seek to minimize the distance 
\(
D(\Lambda^{\mathrm{loc}}(\rho^{\mathrm{int}}_{AB}),\,\rho^{t}_{AB}\ ;M_i)
\)
over all admissible local channels. Substituting the expressions from \eqref{expected_value_local} and \eqref{metric} yields the following cost function on the complex Stiefel manifold:
\begin{align}
J_D^{\text{loc}}(\tilde{S}) &\coloneqq D(\Lambda^{\text{loc}}(\rho^{\mathrm{int}}_{AB}), \rho^t_{AB} \ ; M_i)  \label{distance_function} \\ &= \sqrt{ \sum_{i=1}^{N^2-1} \left( \text{Tr} \left[ \tilde{S} \rho^{\mathrm{int}}_{AB} \, \tilde{S}^\dagger \, U_\sigma^\dagger \tilde{M}_i U_\sigma \right] - \text{Tr} \left[ \rho^t_{AB} \, M_i \right] \right)^2 } \nonumber 
\end{align}
where the definition of $\tilde{M}_i$ is given in Eq.~\eqref{tilde_operators}.
Applying the chain rule to compute the gradient \(\nabla J_{D}^{\mathrm{loc}}\) and then projecting via \eqref{projection_local_1} and \eqref{projection_local_2} is straightforward:
\begin{align}
\nabla J_{D}^{\mathrm{loc}}(\tilde S)
  = \frac{1}{\sqrt{ J_D^{\text{loc}} }} \sum_{i=1}^{N^{2}-1}
     \,\nabla J_{i}^{\mathrm{loc}}\,
     \bigl(
        J_{i}^{\mathrm{loc}}
        - \operatorname{Tr}\!\bigl[\rho^t_{AB}\,M_i\bigr]
     \bigr),
     \label{eq:grad-JDloc} \\[6pt]
\text{with}\qquad
J_{i}^{\mathrm{loc}}(\tilde S)
  = \operatorname{Tr}\!\left[
        \tilde S \rho^{\mathrm{int}}_{AB} \, \tilde S^{\dagger} \,
        U_{\sigma}^{\dagger} 
        \tilde M_i 
        U_{\sigma}
     \right].
     \nonumber
\end{align}

Upon deriving the explicit gradient components, we proceed with the iterative steps introduced from Eq.~\eqref{projection_local_1} to Eq.~\eqref{evolution_local} to minimize the cost function $J_{D}^{\mathrm{loc}}(\tilde S)$.

However, we emphasize that \(J_{D}^{\mathrm{loc}}\) is non-convex. This non-convexity is not solely due to the outer sum in \(J_{D}^{\mathrm{loc}}\), but already exists in the original objective function given in Eq.~\eqref{expected_value_local}. Therefore, different random initializations on the manifold may converge to different local minima.

\subsection{Probabilistic Local Transformation}
We may interpret any CPTP map
\(\Lambda(\rho)=\sum_{i}K_i\rho K_i^{\dagger}\) as a generalized measurement on the
state~\(\rho\).  The Kraus operators \(\{K_i\}\) define a generalized quantum measurement whose POVM elements \(E_i = K_i^{\dagger}K_i\) satisfy \(\sum_i E_i = \mathbb{I}\).  Conditional
on obtaining outcome~\(i\), the post-measurement state is 
\begin{align}
  \rho_i 
  \;=\;
  \frac{K_i\,\rho\,K_i^{\dagger}}
       {\operatorname{Tr}\!\bigl[E_i\,\rho\bigr]},
  \qquad
  \text{with probability }
  \operatorname{Tr}\!\bigl[E_i\,\rho\bigr].\end{align}

In the case of a local Kraus operation in a bipartite system, Alice and Bob can perform their local measurements and then use classical communication to decide whether to keep the post‐measurement state
or to discard it, conditioned on the outcome.\footnote{To avoid misinterpretation, by classical communication we refer solely to the process of deciding whether to keep or discard the output state, with the quantum operations on the two subsystems remaining independent of each other. Consequently, this approach does not cover LOCC.}  Equivalently, Alice and Bob can implement the local generalized measurement and post-selection by coupling their subsystems to a local ancilla, apply a joint local unitary on system plus ancilla, and then measure the ancilla qubits, which is called \emph{Von Neumann indirect measurement}. By exchanging the resulting measurement outcomes, they determine whether to retain or discard the corresponding system qubits.

To incorporate post-selection into the Stiefel manifold parameterization, we right-multiply an \( n \times l \) complex Stiefel matrix \( S \) by a diagonal selector matrix \( \Omega \in \mathbb{C}^{n \times n} \), whose entries are either zeros or ones. This operation effectively keeps only the desired Kraus operators. For instance, in a single-qubit system,
\(
  \Omega_1 = \operatorname{diag}(\mathbb{I}_2, \mathbf{0}_2, \mathbf{0}_2, \mathbf{0}_2)
\)
retains the first \( 2 \times 2 \) block of \( S \in V_2(\mathbb{C}^8) \), yielding
\begin{align}
  \Omega_1 S
  \;=\;
  \begin{pmatrix}
    K_1 \\[2pt]
    \mathbf{0}_2 \\[2pt]
    \mathbf{0}_2 \\[2pt]
    \mathbf{0}_2
  \end{pmatrix}_{8\times 2},
\end{align}
where the matrix \( \Omega_1 \) ensures that only the action associated with the first Kraus operator \( K_1 \) is preserved.

Since the post-selected state must be properly normalized, we divide by the success probability \(\text{Tr}[\Omega\,S\,\rho\,S^\dagger\,\Omega^\dagger]\), which corresponds to the trace of the unnormalized post-selected state. Thus, the original objective function in Eq.~\eqref{expected_value} becomes
\begin{equation}
    J_{C}(S)
    \;\coloneqq\;
    \frac{
      \text{Tr}\ \!\bigl[\Omega S \rho S^{\dagger} \Omega^{\dagger} \,\tilde{\mathcal{O}}\bigr]
    }{
      \text{Tr}\ \!\bigl[\Omega S \rho S^{\dagger} \Omega^{\dagger}\bigr]
    }.
    \label{eq:J_C}
\end{equation}
We name the numerator and denominator traces as
\begin{align}
  J_1(S) \coloneqq \text{Tr}\ \!\bigl[S \rho S^{\dagger} \,\Omega^{\dagger} \tilde{\mathcal{O}} \Omega\bigr],
  \qquad
  J_2(S) \coloneqq \text{Tr}\ \!\bigl[S\rho S^{\dagger} \, \Omega^{\dagger} \Omega\bigr],
\end{align}
representing the expectation value of the observable \(\Omega^{\dagger}\tilde{\mathcal{O}}\Omega\) and \(\Omega^{\dagger}\Omega\), respectively. By applying the chain rule, the gradient reads
\begin{equation}
    \nabla J_{C}(S)
    = 
    \frac{\nabla J_1(S)}{J_2(S)}
    \;-\;
    \frac{J_1(S)\,\nabla J_2(S)}{\bigl[J_2(S)\bigr]^2}.
    \label{eq:grad-JC}
\end{equation}

To impose the locality constraint, we simply replace \(S\) by
\(U_{\sigma}\tilde{S}\) in Eq.~\eqref{eq:J_C}, and after calculating the gradient, we use gradient projection in~\eqref{projection_local_1} and \eqref{projection_local_2}, then repeat the same procedure as nonlocal case:
\begin{align}
J^{\text{loc}}_C(\tilde{S}) \coloneqq \frac{\text{Tr} \left[ \tilde{S} \rho_{AB} \, \tilde{S}^\dagger \, U_\sigma^\dagger \Omega^\dagger \tilde{\mathcal{O}} \Omega U_\sigma \right]}{\text{Tr} \left[ \tilde{S} \rho_{AB} \, \tilde{S}^\dagger \, U_\sigma^\dagger \Omega^\dagger \Omega U_\sigma \right]} , \label{J_loc}
\end{align}
where the corresponding success probability is given by
\begin{align}
    P_\text{succ}=\text{Tr} \left[ \tilde{S} \rho_{AB} \, \tilde{S}^\dagger \, U_\sigma^\dagger \Omega^\dagger \Omega U_\sigma \right] \label{Eq. probablity of success}
\end{align}

This extension to probabilistic local transformation finds an application in the optimization of entanglement distillation protocols, where non-trace-preserving maps are employed to probabilistically increase the fidelity of a shared 2-qubit state with Bell states.

\section{Application: Entanglement Distillation for Low-FEF States} \label{4}
In this section, we apply the introduced numerical approachs to the entanglement distillation problem, focusing on distilling weakly entangled states that cannot be distilled using conventional recurrence protocols. We consider two strategies, as detailed below.

In the first approach (\hyperref[Preprocessing via State Transformation]{1}), we aim to transform the initial state into a specific form, referred to as \emph{R-states} in Ref.~\cite{rozpkedek2018optimizing}, which are suitable for the \emph{extreme photon loss} (EPL) protocol, a one-round entanglement distillation scheme \cite{campbell2008measurement}. Specifically, we demonstrate that by adding a tailored pre-processing step into the EPL distillation protocol, it becomes possible to distill entangled states whose fully entangled fraction (FEF) is below \( 0.5\), a regime in which conventional recurrence protocols typically fail. To illustrate the effectiveness of the process, we apply the proposed operation to randomly generated weakly entangled states and show the corresponding plot in the results section (Fig.~\ref{fig:FEF Histogram}). 

In the second approach (\hyperref[Probabilistic Local POVM]{2}), we apply an optimized set of generalized local measurements to a 2-qubit system such that, after post-selection on the local measurements' outcomes, the fidelity with the Bell state \(|\Psi^+\rangle\langle\Psi^+|\) is maximized. This method can be regarded as an extension of \emph{filtering protocol} introduced in Ref.~\cite{gisin1996hidden}, where the two subsystems apply fixed generalized quantum measurement operators to a set of noisy entangled states to distill higher entanglement. In contrast, our approach tailors the generalized measurement operators to the known initial state, thereby enhancing the output fidelity and enabling a more efficient distillation process. In the Results section, we present the performance of this method on weakly entangled states and provide a comparison of the output fidelity of this optimization method with the upper bound for distilled fidelity given in Ref.~\cite{rozpkedek2018optimizing}.

\subsection{Approach 1: Preprocessing Through Local State Transformation} \label{Preprocessing via State Transformation}

\subsubsection*{Methodology} \label{Methodology - Ap1}
One of the optimized distillation protocols for a specific initial state, referred to as the R-state, is the Extreme Photon Loss (EPL) protocol \cite{campbell2008measurement}. The R-state is a mixture of a Bell state \(|\Psi^\pm \rangle=(|01\rangle\pm|10\rangle)/\sqrt{2}\) and separable noise \(|11\rangle\), given by
\begin{align}
\rho_R(p) = p |\Psi^\pm \rangle \langle \Psi^\pm| + (1 - p) |11\rangle \langle 11|. \label{rho_R}
\end{align}

The EPL distillation protocol uses two pairs of R-states as input to probabilistically produce one pair of qubits in a maximally entangled Bell state, with a success probability of \(P_{\text{succ}} = \frac{p^{2}}{2}\). Importantly, the EPL protocol is optimal as it produces the maximally entangled Bell state with the highest possible probability when the input is an R‑state \cite{rozpkedek2018optimizing}.

\begin{figure}
    \centering
    \[
    \begin{quantikz}[row sep=0.7cm, column sep=0.7cm]
        \lstick{$A_1$} && \gate{\Lambda_A} &&\ctrl{1}  &&  \\
        \lstick{$A_2$} && \gate{\Lambda_A} &&\targ{} &  \meter{\sigma_z}
        \\ 
        \lstick{$B_1$}   && \gate{\Lambda_B} && \ctrl{1} && \\
        \lstick{$B_2$}   && \gate{\Lambda_B} && \targ{}  & \meter{\sigma_z} 
    \end{quantikz}
    \]
    \caption{First, we transform each pair of qubits from state $\rho^{\mathrm{int}}_{AB}$ to state $\rho^\prime_{AB} $ using the mapping $\Lambda_A \otimes \Lambda_B$. These mappings are chosen such that the state $\rho^\prime_{AB} $ closely approximates an R-state. Then, as depicted, we apply the EPL protocol, CNOT operation followed by measurement, on two pairs of qubits $\rho'_{AB} \otimes \rho'_{AB}$.}
    \label{fig:EPL_curcuit} 
\end{figure}
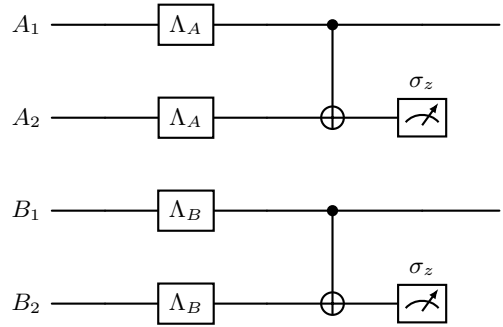

In this setting, we seek a local quantum channel \(\Lambda^{\text{loc}}=\Lambda_A\otimes\Lambda_B\) that transforms—exactly or approximately—a given 2-qubit state \(\rho^{\mathrm{int}}_{AB}\) into a target state $\rho^t_{AB}$, in the form of an R-state. Subsequently, two copies of the transformed state \( \rho'_{AB}=\Lambda_A\otimes\Lambda_B(\rho^{\mathrm{int}}_{AB}) \) can then be processed by the EPL protocol to distill entanglement, as depicted in Fig.~\ref{fig:EPL_curcuit}.

One of the most significant advantages of employing this method lies in its ability to distill entangled states with a fully entangled fraction less than or equal to 0.5, which cannot be distilled using conventional distillation protocols, such as the BBPSSW and DEJMPS protocols. In contrast, the EPL protocol requires the initial state to be an R-state with parameter \( p > 0 \) and can distill states with \(\mathrm{FEF}<0.5\). Consequently, by approximately transforming a given initial state into an R-state, this approach enables the distillation of states whose fully entangled fraction does not exceed 0.5, thus overcoming a limitation associated with standard distillation protocols.

\subsubsection*{Numerical results} \label{results - Ap1}

\begin{figure}
    \centering
    \includegraphics[width=1\linewidth]{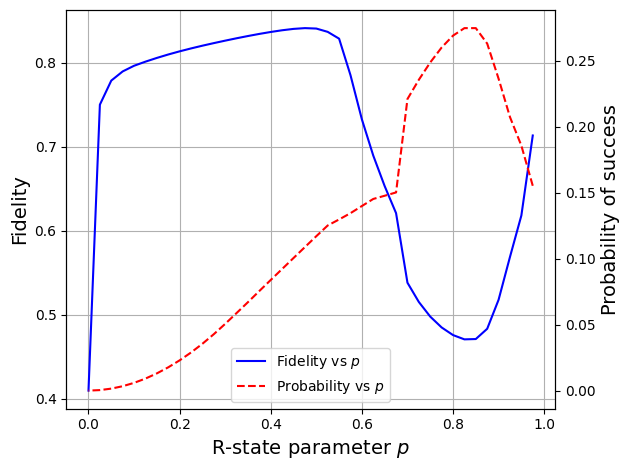}
    \caption{Output fidelity and the corresponding success probability plotted as functions of the parameter \(p\) in the R-state, which serves as the target state. For each value of \(p\), the initial state \(\rho_{AB}^{*}\) is first transformed via an optimized channel to approximate the target R-state \(\rho_{R}(p)\). The EPL distillation protocol is then applied to the transformed state, and both the fidelity of the resulting state and the probability of success are subsequently calculated.}
    \label{fig:fidelity-p plot}
\end{figure}

To parameterize the target R-state, it is necessary to fix the coefficient \(p\) in Eq.~\eqref{rho_R}. However, there is no deterministic prescription for selecting the value of \(p\), nor for expressing \(p\) as a direct function of a given initial density matrix \(\rho^{\mathrm{int}}_{AB}\). As a result, we proceed by iterating the optimization procedure over a series of possible values for the coefficient \(\{0<p_i<1\}\). For each value \(p_i\), we identify the optimized channel, apply the EPL protocol to the resulting state \(\rho'_{AB}\), and subsequently evaluate the fidelity of the output state with the Bell state (see Fig.~\ref{fig:fidelity-p plot}). 

For the purpose of this optimization, since the target state is an R-state, the set of corresponding observable operators \(\{M_i\}\) is the set
\(
  \bigl\{\,|11\rangle\langle 11|,\
           |\Psi^{+}\rangle\langle \Psi^{+}|,\
           i\bigl(|\Psi^{+}\rangle\langle 11| - |11\rangle\langle \Psi^{+}|\bigr)/\sqrt{2},\
           \bigl(|\Psi^{+}\rangle\langle 11| + |11\rangle\langle \Psi^{+}|\bigl)/\sqrt{2}\bigr\}.
\)

Now, let us consider the following state as an example to illustrate the optimization procedure:
\begin{align}
\rho_{AB}^\ast = \begin{pmatrix}
 0.3   & -0.11 &  0.19 &  0.14 \\
-0.11  &  0.24 & -0.11 &  0.15 \\
 0.19  & -0.11 &  0.14 &  0.03 \\
 0.14  &  0.15 &  0.03 &  0.32
\end{pmatrix}.
\end{align}
This state is entangled, as quantified by its concurrence~\cite{hill1997entanglement}, \(\mathcal{C}(\rho_{AB}^\ast) = 0.2\). 
However, its FEF lies below the threshold of \(0.5\) (\(\mathrm{FEF}(\rho_{AB}^\ast)\approx 0.489\)) required for distillation using the BBPSSW and DEJMPS protocols. Moreover, since this state does not have the form of an R-state, it does not satisfy the requirement for distillation via the EPL protocol.

As previously described, we consider multiple values of the parameter \( p \in [0,1] \). For each chosen value of \( p \), we first determine an optimized quantum channel that transforms the initial state \(\rho_{AB}^*\) as closely as possible into the corresponding R-state \(\rho_R(p)\). Following this optimization, we apply the EPL distillation protocol to the resulting state and subsequently plot the fidelity of the output state with the Bell state \(|\Psi^+\rangle\) as a function of the parameter \( p \). As shown in Fig.~\ref{fig:fidelity-p plot}, there exist specific values of \( p \) for which the fidelity of the output state, after performing the EPL protocol, attains its maximum value. Notably, for these particular values of \( p \), the achieved fidelity exceeds 0.5, making it possible to employ conventional protocols as subsequent distillation steps on the output state of this procedure. This
demonstrates the effectiveness of the protocol in producing entangled states.

\begin{figure}
    \centering
    \includegraphics[width=1\linewidth]{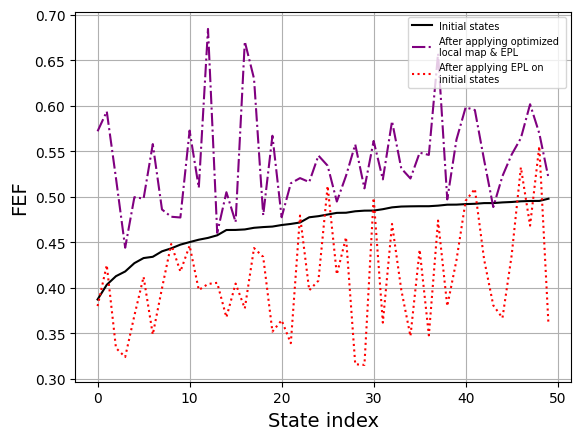}
    \caption{Comparison of the FEF for 50 randomly generated 2-qubit states before and after processing. The black solid line represents the FEF of the initial states, the red dotted line shows the FEF after applying the EPL protocol directly to the initial states, and the purple dash–dotted line corresponds to the FEF obtained after applying the optimized local maps followed by the EPL protocol. As shown, approximately \(75\%\) of the processed states achieve an FEF exceeding \(0.5\).}
    \label{fig:FEF Histogram}
\end{figure}

In the next step, we apply the same procedure to 50 randomly generated entangled states with \(\mathrm{FEF} < 0.5\).\footnote{The random density matrices are drawn from a uniform distribution using the Wishart ensemble, and only those that are entangled and satisfy \(\mathrm{FEF} < 0.5\) are retained.} For each generated state, we attempt to transform it as closely as possible into an R-state with a parameter \(p\) that yields the highest fidelity, and then apply the EPL protocol, subsequently measuring the FEF of the resulting states. 
Figure~\ref{fig:FEF Histogram} compares the FEF of three states: (i) initial states, (ii) after applying the EPL protocol directly to the initial state, and (iii) after first transforming the initial states using the optimized local map and then applying the EPL protocol. As shown, approximately \(75\%\) of the initial states achieve an FEF exceeding \(0.5\) after transformation and distillation.

\subsection{Approach 2: Entanglement Distillation Using Probabilistic Local Transformation} \label{Probabilistic Local POVM}

\subsubsection*{Methodology} \label{Methodology - Ap2}

This setting closely resembles the filtering protocol~\cite{gisin1996hidden} in terms of the number of qubits processed (\(1 \to 1\) distillation protocol). Specifically, in this approach, each qubit is subjected to a local generalized measurement with four measurement operators. The post-measurement state is retained only when Alice and Bob obtain the same specific measurement outcome from the set \(\{0,1,2,3\}^2\), in particular the outcome \(00\); otherwise, it is discarded.

Our goal is to optimize these measurement operators. Concretely, given an input 2-qubit state \(\rho_{AB}^{\mathrm{int}}\), we seek sets of Kraus operators \(\{K_i^A\}_{i=1}^4\) and \(\{K_j^B\}_{j=1}^{4}\) that maximize the fidelity of the output state with \( |\Psi^+\rangle\) in the case of successful measurements. In Eq.~\eqref{J_loc}, we defined the optimization objective for this approach, setting \(\mathcal{O} = |\Psi^+\rangle \langle \Psi^+|\) (\(\tilde{\mathcal{O}} = \mathbb{I}_{16} \otimes |\Psi^+\rangle \langle \Psi^+|\)), and \(\Omega = \mathrm{diag}(\mathbb{I}_4,\mathbf{0}_4,\ldots,\mathbf{0}_4)_{64} \).

\subsubsection*{Numerical results} \label{results - Ap2}

We now evaluate the performance of the probabilistic local transformation method. First, we compare its output fidelity with the theoretical upper bound for a specific example. Next, we extend the analysis to 50 random samples of weakly entangled real states, again comparing the optimized fidelity to its corresponding upper bound. Finally, we compare the performance of the probabilistic method against the first approach (preprocessing via state transformation) by applying both to the same set of samples from the previous example (Fig.~\ref{fig:FEF Histogram}) and evaluating the resulting fidelities.

\begin{figure}
    \centering
    \includegraphics[width=1\linewidth]{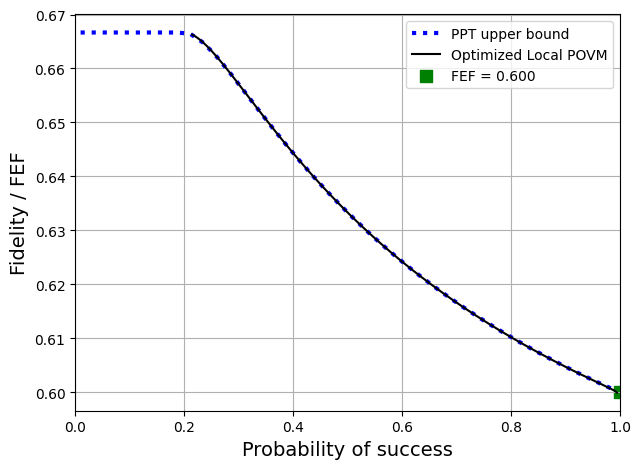}
    \caption{Fidelity versus success probability for the initial state \(\rho_{S}(0.2)\). The black solid curve shows the optimization trajectory on the Stiefel manifold, while the blue dotted line represents the PPT upper bound. The optimized curve closely follows the bound, indicating that the protocol effectively saturates the theoretical limit.}
    \label{fig:plot fidelity vs probability Filtering}
\end{figure}

The upper bound on the distilled fidelity is obtained by optimizing over the set of positive partial transpose (PPT) operations\footnote{PPT operations are quantum channels whose corresponding Choi matrix satisfies the positive partial transpose (PPT) criterion.}, which strictly contains the set of LOCC operations but is not necessarily implementable~\cite{rains1999bound}. For a given success probability, this bound can be efficiently computed via semidefinite programming (SDP) \cite{rozpkedek2018optimizing}, as shown in Fig.~\ref{fig:plot fidelity vs probability Filtering}. 

We first analyze the performance of our method on a single representative state. Specifically, we consider the family of states
\begin{align}
    \rho_{S}(p) = p\,\bigl|\Psi^{+}\bigr\rangle\!\bigl\langle \Psi^{+}\bigr| 
    + (1 - p)\ |01\bigr\rangle\!\bigl\langle 01\bigr| ,
\end{align}
which are mixtures of a maximally entangled Bell state and non-orthogonal noise components.

\begin{figure}
    \centering
    \includegraphics[width=0.6\linewidth]{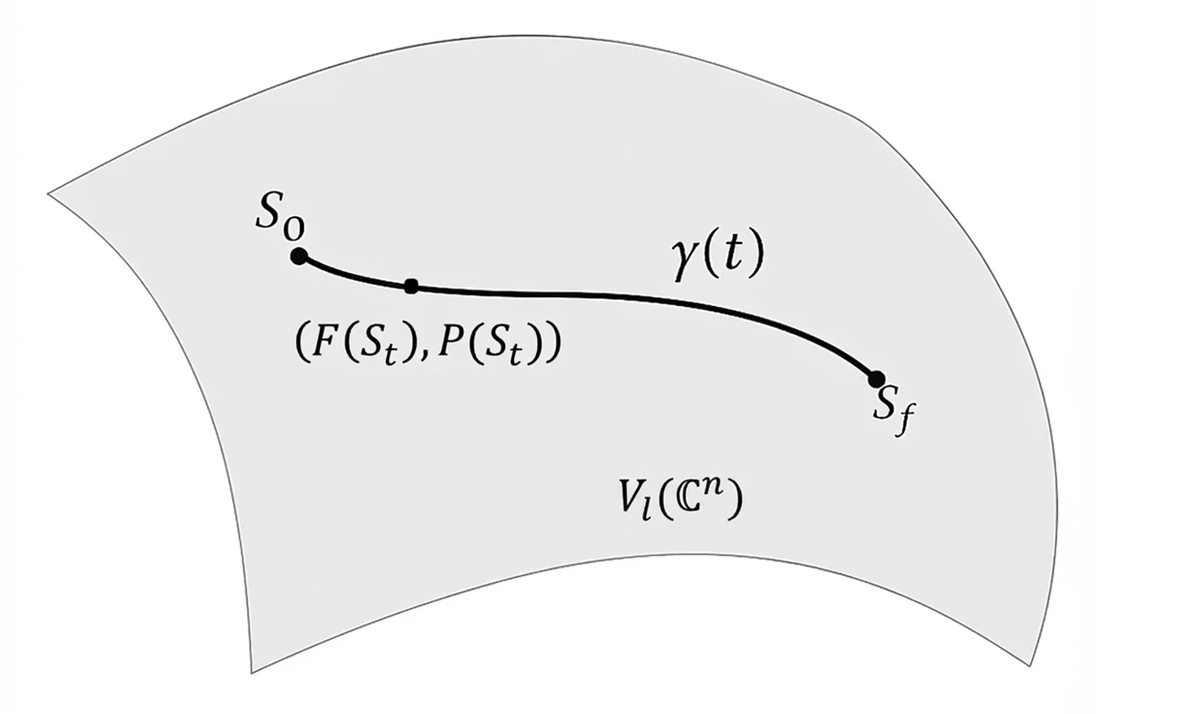}
    \caption{Starting from the point \(S_0\) on the manifold \(V_l(\mathbb{C}^{n})\), we follow the trajectory \(\gamma(t)\) to the optimal point \(S_f\) and evaluate the output fidelity \(F(S_t)\) and success probability \(P(S_t)\) for each intermediate channel associated with the points \(S_t\) along the path.}
    \label{fig:stiefel-trajectory}
\end{figure}

We apply the optimization procedure to a member of \(\rho_{S}\) with parameter \(p = 0.2\), for which \( \mathrm{FEF}\big(\rho_S(0.2)\big)=0.6 \). As demonstrated in Fig.~\ref{fig:stiefel-trajectory}, starting from an initial point on the Stiefel manifold associated with a deterministic fidelity maximizing channel\footnote{Such channels can be obtained during the calculation of the fully entangled fraction.}, denoted \(S_0\), we apply a gradient ascent algorithm on the manifold to trace a continuous trajectory \(\gamma(t)\) from \(S_0\) to the optimal point \(S_f\). Each intermediate point \(S_t\) along this trajectory corresponds to a candidate channel, for which we evaluate the post-selected fidelity \(F(S_t)\) and the associated success probability \(P(S_t)\), as defined in Eq.~\eqref{J_loc} and Eq.~\eqref{Eq. probablity of success}. The resulting fidelity–probability trade-off is plotted in Fig.~\ref{fig:plot fidelity vs probability Filtering}, indicating that the fidelity-probability curve coincides the upper bound.

\begin{figure}
    \centering
    \includegraphics[width=1\linewidth]{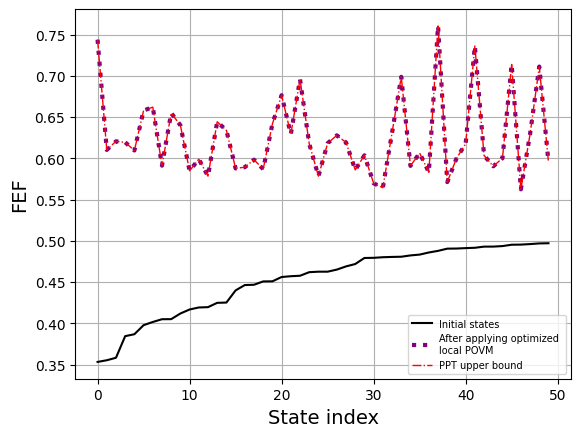}
    \caption{Comparison of the FEF for 50 randomly generated real density matrices. The black solid line shows the initial FEF values (all below \(0.5\)). The purple dotted line and the red dash-dotted line, which overlap, represent the fidelity of the post-selected states after applying the optimized probabilistic local operations and the upper bound, respectively. The complete overlap indicates that the protocol achieves the theoretical limit for all tested states.}
    \label{FEF_state index (2nd approach)}
\end{figure}

We extend the analysis to an ensemble of \(50\) randomly generated entangled \emph{real} density matrices in a 4-dimensional Hilbert space, each with an initial FEF strictly less than \(0.5\).\footnote{Real density matrices are chosen since the upper bound can be computed for them. The theoretical upper bound is computed using semidefinite programming, which imposes restrictions for complex-valued matrices.} For all test states, the fidelity of the post-selected states coincides with the upper bound, confirming the optimality of the protocol. The results are summarized in Fig.~\ref{FEF_state index (2nd approach)}.

\begin{figure}
    \centering
    \includegraphics[width=1\linewidth]{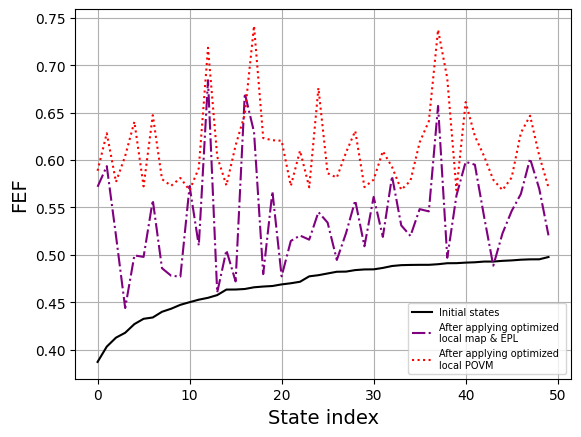}
    \caption{Comparison of the increase in FEF for 50 randomly generated two-qubit states using the two proposed approaches. The black solid line represents the initial FEF values (all below \(0.5\)). The red dotted and purple dash-dotted lines represent, respectively, the FEF of the post-selected states after applying the optimized probabilistic local operation and the FEF of the distilled states obtained by applying the optimized local map followed by the EPL protocol. As illustrated, the FEF values generally increase for almost all sampled states when directly optimizing the probabilistic local operations.}
    \label{fig:placeholder}
\end{figure}

Finally, we compare the performance of the two approaches: state transformation as a preprocessing step and probabilistic local transformation. Both methods were applied to the same set of 50 randomly generated input states of Fig.~\ref{fig:FEF Histogram}, and a comparison of the resulting FEFs is presented in Fig.~\ref{fig:placeholder}. We observe that the probabilistic approach consistently outperforms the preprocessing method.
This result is expected, as the second method corresponds to a direct fidelity optimization, while the first one relies on an indirect optimization. Both methods achieve comparable success probabilities in the range of \(0.1\)–\(0.2\).

\section{Summary and Future Works} \label{5}

In this work, we have introduced a numerical framework to optimize local CPTP maps for transforming an arbitrary bipartite mixed state into a target state both deterministically and probabilistically, leveraging a complex Stiefel manifold parameterization of Kraus operators. By formulating the task as a gradient-based minimization of a cost function, our method overcomes the lack of analytic criteria for mixed state convertibility under locality constraints. A key contribution lies in the handling separability constraints, where we ensure the locality of the channels while efficiently navigating the complex Stiefel manifold.

We demonstrated two substantive applications in entanglement distillation. 
First, by pre-processing arbitrary weakly entangled inputs to optimally approximate a target R-state, we extend the applicability of the EPL protocol, thereby making it effective for any 2-qubit state. This generalization includes states with a fully entangled fraction (FEF) below \(0.5\), for which conventional recurrence protocols fail.
Second, we optimized the local generalized measurements for probabilistic transformation to a target Bell state by directly maximizing the fidelity. Our numerical simulations demonstrate that the achieved fidelities saturate the known theoretical upper bound. Furthermore, we provide numerical examples which confirm that the direct optimization approach exhibits superior performance compared to the pre-processing approach. 
  
Beyond entanglement distillation, our framework is applicable to other quantum control tasks requiring locality, such as distributed state discrimination, and quantum network routing. The Stiefel manifold-based approach can be generalized to problems with locality constraints in multipartite systems, enabling the optimization of local operations across arbitrary quantum networks.

\section*{Code Availability}
The code used to generate all results and figures in this study is openly available at  
\url{https://github.com/SasanSarbishegi/Optimizing-Quantum-State-Transformation}\,.

\section*{Acknowledgment}
This research was supported in part by Iran National Science Foundation,
under Grant No.4022322.

\bibliographystyle{unsrt}
\bibliography{Bib}

\end{document}